\pdfoutput=1
\documentclass[aps,pra,reprint,amsmath,amssymb,superscriptaddress]{revtex4-1}
\usepackage{bm}
\usepackage{graphicx}
\usepackage[colorlinks]{hyperref}
\usepackage{mathrsfs}
\usepackage{times}

\begin{document}

\title{Surface waves and bulk Ruderman mode of a bosonic superfluid vortex crystal in the lowest Landau level}

\author{Bhilahari Jeevanesan}
\affiliation{Physik-Department, Technische Universit\"at M\"unchen, 85748 Garching, Germany}
\affiliation{Munich Center for Quantum Science and Technology (MCQST), 80799 M\"unchen, Germany}
\author{Claudio Benzoni}
\affiliation{Physik-Department, Technische Universit\"at M\"unchen, 85748 Garching, Germany}
\affiliation{Munich Center for Quantum Science and Technology (MCQST), 80799 M\"unchen, Germany}
\author{Sergej Moroz}
\affiliation{Physik-Department, Technische Universit\"at M\"unchen, 85748 Garching, Germany}
\affiliation{Munich Center for Quantum Science and Technology (MCQST), 80799 M\"unchen, Germany}
\affiliation{Department of Engineering and Physics, Karlstad University, Karlstad, Sweden}

\begin{abstract}
We determine and analyze collective normal modes of a finite disk-shaped two-dimensional vortex crystal formed in a compressible bosonic superfluid in an artificial magnetic field. Using the microscopic Gross-Pitaevskii theory in the lowest Landau level approximation, we generate vortex crystal ground states and solve the Bogoliubov-de Gennes equations for small amplitude collective oscillations. We find chiral surface waves that propagate at frequencies larger than those of the bulk Tkachenko modes. Furthermore, we study low frequency bulk excitations and identify a Ruderman mode, which we find is well-described by a previously developed low-energy effective field theory. 

\end{abstract}

\maketitle


\section{Introduction}
\label{sec:intro}

Quantum vortex crystals, first predicted by Abrikosov for type II superconductors in an external magnetic field \cite{Abrikosov1957}, emerge also in neutral superfluids realized in Helium and cold atoms experiments \cite{Cooper2008, Fetter2009, RevModPhys.80.885, Sonin2016, Saarikoski2010}. In these cases the background magnetic field can be mimicked by external rotation or with other types of artificial gauge fields \cite{aidelsburger2018artificial}. As a result of such a background field, time-reversal symmetry is explicitly broken, which manifests itself in chiral elliptically-polarized collective oscillations, known as Tkachenko waves \cite{sonin2014tkachenko}.

In recent investigations \cite{benzoni2021rayleigh, marijanovic2021rayleigh} we explored an effective field theory (EFT) of two-dimensional elastic media with Lorentz-type forces that break time-reversal symmetry. Within this approach, which provides a suitable description of skyrmion crystals in thin-film chiral magnets and gyroscopic metamaterials, we established the existence of Rayleigh surface waves with an unusual property: The chirality of the Rayleigh waves is fixed not only by the sign of the effective magnetic field, but is also determined by the elasticity properties of the crystal. In the simplest case of a crystal with triangular symmetry one finds three different regions which differ by the chirality of the Rayleigh waves when the Poisson ratio is varied. 

\begin{figure}[t] 
\centering{}\includegraphics[width= 5.5 cm]{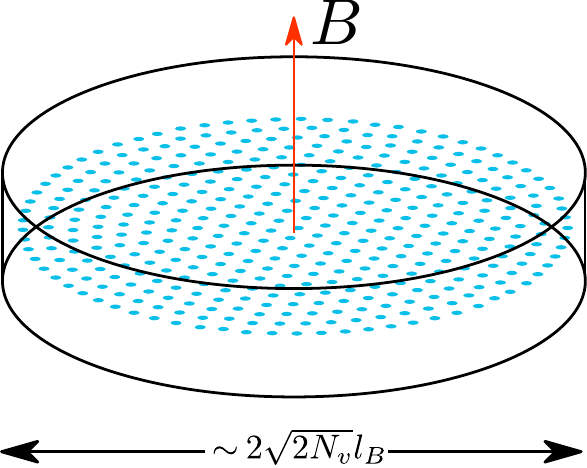}\protect\caption{We consider a droplet of bosons that experience a synthetic magnetic field $B$ and can occupy only the lowest $N_v+1$ angular momentum LLL orbitals. The mean-field LLL wavefunction is specified by the position of the $N_v$ vortices (blue). The bosons interact via a short-range repulsive contact interaction, resulting in a vortex crystal ground state.  \label{fig:Intro} }
\end{figure}

As mentioned above, time-reversal symmetry is explicitly broken in a superfluid vortex crystal. Hence, it is very natural to ask if this quantum state of matter, where elasticity and superfluid low-energy degrees of freedom interact with each other, also supports chiral elastic Rayleigh waves. This question can be addressed within the framework of a low-energy EFT for vortex lattices, developed in \cite{redundancies, effectivevortex, moroz2019bosonic, nguyen2020fracton}, see also \cite{Sonin1976, volovik1979,  Baym2003} for related preceding works on vortex crystal hydrodynamics. We find in appendix \ref{AppA}  that within the leading order EFT of two-dimensional superfluid vortex crystals, low-frequency Rayleigh waves are absent. For this reason, we go beyond the low-energy realm in this paper and  address the question of chiral edge excitations by analyzing the collective oscillations of superfluid vortex crystals within the microscopic Gross-Pitaevskii framework in the geometry illustrated in Fig.\ref{fig:Intro}.

In the two-dimensional setting, studies of rotating superfluids with a finite number of vortices have been undertaken in the past, starting with \cite{campbell1979vortex, campbell1981edge}, for a recent review see  \cite{Sonin2016}. These studies analyze rotating superfluids in the incompressible regime, where vortices interact with each other through a logarithmic two-body interaction. In this regime, edge waves of different types were identified and recently investigated in \cite{PhysRevLett.122.214505, cazalilla2003surface, patil2021chiral, hocking2022anti}. Beyond the incompressible limit, the edge collective modes of arrays of vortices in superfluids have been analyzed previously in \cite{baksmaty2005chiral}.

In the present study we investigate bulk and edge collective modes of a bosonic superfluid confined in a circular geometry and stabilized by short-range repulsive interactions between bosons. To incorporate the effects of an artificial magnetic field, we work in the lowest Landau level (LLL) regime by expanding the condensate wave-function in terms of only the LLL basis-functions.  Previously, this approach was employed for example in \cite{butts1999predicted, PhysRevA.70.033604, Kavoulakis2000, aftalion2005vortex} to determine vortex crystal ground states of rotating cold atom superfluids under harmonic confinement and also in a study of LLL thermodynamics \cite{jeevanesan2020thermodynamics}. The interest in the LLL superfluid physics was rekindled by recent experimental works: In \cite{chalopin2020probing} the LLL limit is reached through a striking use of a synthetic dimension in the form of atomic spin. Another remarkable work \cite{fletcher2021geometric} showed that by a procedure named `geometrical squeezing' by the authors, a gas of sodium atoms can be brought into a superfluid LLL state. Spontaneous crystalization of the resulting LLL condensate has been observed in \cite{mukherjee2022crystallization}.

The advantage of studying the finite vortex crystal in the LLL limit lies in the simplicity of the simulation: In the LLL regime, the Gross-Pitaevskii equation reduces from a partial differential equation in two dimensions, eqs. \eqref{eq:GP} and \eqref{eq:EnergyFunctional}, to a coupled non-linear system of $N_v+1$ ordinary differential equations (ODE) \eqref{eq:GPforCoeffs} \cite{biasi2017exact}, where $N_v$ is the number of quantum vortices in the superfluid droplet \footnote{ The physical reason for the resulting simplicity lies in the fact that for a fixed number of particles the LLL condensate wavefunction is completely determined by the location of its zeros. Note, however, that in the LLL regime the interaction between vortices turns out to be of a multibody nature and thus cannot be decomposed into a sum of two-body potentials \cite{bourne2007vortices}}. 

In this paper, we linearize the LLL Gross-Pitaevskii equations around the vortex crystal ground state and work out the small-amplitude oscillation spectrum illustrated in Fig. \ref{fig:FrequencySpectrum}. We find that it is comprised of low-energy bulk excitations and high-energy chiral surface modes. First, at low energies we obtain an analytical understanding of the nature of collective modes by employing the vortex lattice EFT developed in \cite{redundancies, effectivevortex, moroz2019bosonic, nguyen2020fracton}. In the disk geometry, one of the lowest finite-frequency eigenmodes turns out to be identical to the Ruderman mode, see Fig. \ref{fig:TorsModes}, discovered in \cite{ruderman1970long} in an attempt to explain slow oscillations in the periods of pulsars after glitches. 
Next, we find that the high-frequency modes are localized near the boundary and propagate strictly in one direction around the vortex crystal with the chirality controlled by the sign of the magnetic field, see Fig. \ref{fig:SFModes}. We work out the dispersion relation of these modes in Fig. \ref{fig:DispEdge}.


\begin{figure*}[t] 
\centering{}\includegraphics[width= 17.0 cm]{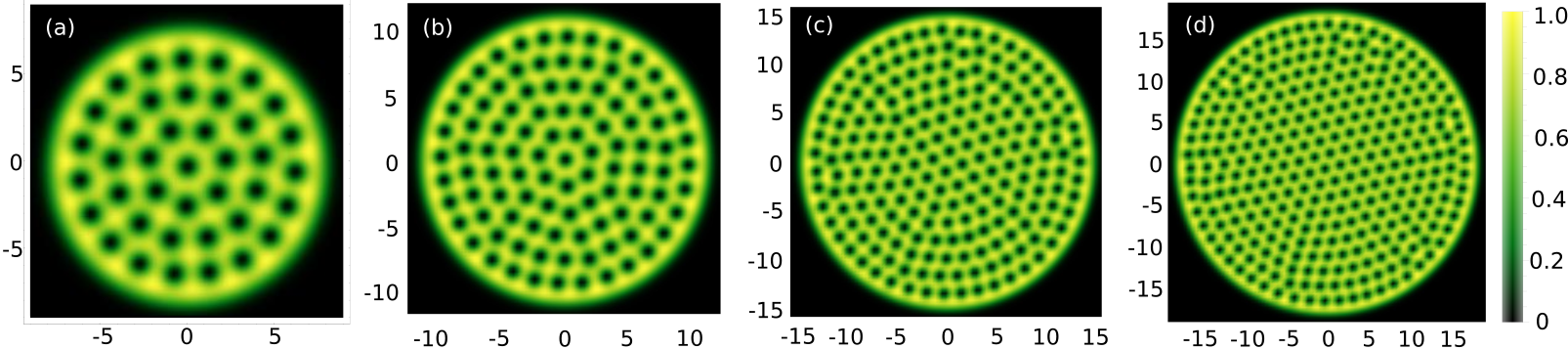}\protect\caption{\label{fig:VortexLattice} Density plots of the ground state wavefunctions of the LLL Gross-Pitaevskii equation for $N_v = 40, 100, 200$ and $300$ vortices. The plots show the normalized boson density $|\psi(x,y)|^2/\max\limits_{x,y} |\psi(x,y)|^2$. Length is measured in units of the magnetic length $l_B=1/\sqrt{B}$.}
\end{figure*}

\section{Gross-Pitaevskii equation and collective oscillations of the vortex lattice in the LLL regime}
\label{sec:oscillations}
The system that we study is a droplet of identical bosons residing in two dimensions and experiencing a constant magnetic field $B$ that acts perpendicularly to the plane, see Fig. \ref{fig:Intro}. Such a magnetic field can also be effectively mimicked by rotating the bosons inside a harmonic trap \cite{ho2001bose}. Alternatively, in recent years advances in the field of ultracold atoms have made it possible to realize artificial magnetic fields through the use of synthetic dimensions \cite{fletcher2021geometric, chalopin2020probing}. 

In the following we consider bosons that interact with each other by means of a contact repulsive potential. Our starting point is the Gross-Pitaevskii equation governing the bosonic condensate, which provides a good description of the Bose gas at vanishing temperature \cite{pethick2008bose, pitaevskii2016bose},  
\begin{eqnarray}
\label{eq:GP}
i \dot{\psi} &=&  \frac{\delta E[\psi,\bar \psi]}{\delta \bar \psi}
\end{eqnarray}
with the energy functional given by
\begin{eqnarray}
\label{eq:EnergyFunctional}
E[\psi,\bar \psi] &=& \int d^2 x \left[\frac{1}{2m} \bar \psi(-i\nabla - {\bf A})^2 \psi +\frac g 2 \left( \bar \psi \psi \right)^2\right],
\end{eqnarray}
where $\bm A$ is the vector potential corresponding to the constant magnetic field $ B$. We choose to work in the symmetric-gauge, for which $\bm A =B/2 (- y, x)$. In this paper we set $\hbar = 1$ and absorb the artificial electric charge of the bosons into the magnetic field. 
We fix the total number of bosons in the system to a value $N$ by requiring
\begin{eqnarray}
\label{eq:ParticleNumber}
N &=& \int d^2 x \bar \psi \psi. 
\end{eqnarray}

The kinetic part of the Gross-Pitaevskii equation has the Landau level eigenstates as solutions. The spacing between these levels is fixed by the cyclotron frequency $\omega_c = B / m$. We now take $m\rightarrow 0$, implying that the cyclotron frequency tends to infinity. This is the LLL limit, since higher Landau levels are infinitely costly and are therefore inaccessible. In the symmetric gauge the LLL eigenstates have the form 
\begin{eqnarray}
\label{eq:basisfunctions}
\psi^{\text{LLL}}_n(z) = \mathcal N_n z^{n} e^{-|z|^2 /4 l_B^2},
\end{eqnarray}
with $z = x + i y$, the magnetic length $l_B=1/\sqrt{B}$ and the normalization constant $\mathcal N_n = {1}/{\sqrt{2^{n+1}\pi n!}l_{B}^{n+1}}$. 

Given that the bosons reside entirely in the LLL, we can simplify the Gross-Pitaevskii equation \eqref{eq:GP} by expanding $\psi(z)$ in terms of the LLL orbitals $\psi^{\text{LLL}}_n(z)$. To this end we insert the ansatz 
\begin{eqnarray} \label{psiLLL}
\psi(z,t)=\sum _{n = 0} ^{N_v} c_n(t) \psi^{\text{LLL}}_n(z)
\end{eqnarray}
into \eqref{eq:ParticleNumber} and \eqref{eq:EnergyFunctional} and carry out the spatial integration. Notice that above we restricted the sum to the $N_v+1$ lowest angular momentum LLL orbitals, which fixes the radius of the droplet roughly to $R=\sqrt{2N_v}l_B$. Physically, this can be achieved by confining the bosons inside a disk of radius $R$ by means of an external, radially symmetric potential. If such a potential is carefully chosen, it will strongly  suppress occupation of all LLL orbitals with $n > N_v$. In appendix \ref{app:Potential} we investigate suitable potentials. In particular, in the first part \ref{app:SoftCutoff} we show that a radial step-function potential approximately realizes such a suppression. In the second part \ref{app:HardCutoff} we derive an exact real-space potential that penalizes the occupation of all orbitals with $n > N_v$ and leaves all orbitals $n\leq N_v$ untouched. 

For the particle number we obtain
\begin{eqnarray}
\label{eq:ParticleNumber_mod}
N = \sum_{n=0}^{N_v} \bar{c}_n c_n,
\end{eqnarray}
while the expression for the energy functional becomes
\begin{eqnarray}
\label{eq:EnergyFunctional_mod}
E[\{c_n,\bar c_n\}] &=& g' \sum_{s = 0}^{2 N_v} \left| \sum_m p_{m}^{s} c_m c_{s-m}\right|^2,\\
p_{m}^{s} &\equiv& \sqrt{2^{-s}{s \choose m}}.
\end{eqnarray}
Here we introduced the shorthand $g' = {g}/{4 \pi l_B^2}$.
The index $m$ in the inner sum is to be extended over $0 \leq m \leq N_v$ with $0\leq s - m \leq N_v$. Above we have dropped the constant  term ${\omega_c N}/{2} $ stemming from the kinetic energy operator in \eqref{eq:EnergyFunctional}. This amounts to measuring the energy from zero at the LLL.

Finally, the Gross-Pitaevskii equations for the LLL amplitudes $c_n$ are obtained by taking the functional derivative of the energy \eqref{eq:EnergyFunctional_mod} with respect to $\bar c_n(t)$
\begin{equation}
\label{eq:GPforCoeffs}
\begin{split}
i \dot c_n &= \frac{ \delta E[\{c_n,\bar c_n\}]}{\delta \bar c_n}   \\
&= 2g'\sum_{s=0}^{2N_{v}}\left(\sum_{m}p_{m}^{s}c_{m}c_{s-m}\right)p_{n}^{s}\bar{c}_{s-n}.
\end{split}
\end{equation}
This coupled set of $N_v+1$ non-linear ODEs is fully equivalent to eqs. \eqref{eq:GP} and \eqref{eq:EnergyFunctional} in the LLL limit. Since eq. \eqref{eq:GP} leads to a PDE in two dimensions, we gain considerable computational simplification by working with eq. \eqref{eq:GPforCoeffs} instead. It is straightforward to check using eqs. \eqref{eq:ParticleNumber_mod} and \eqref{eq:GPforCoeffs} that particle number is preserved under time-evolution.

\subsection{Vortex lattice ground state}
\label{sec:vortexLatticeGS}

As a first step towards studying the collective oscillations of the quantum system, we find its ground state by minimizing the total energy \eqref{eq:EnergyFunctional_mod} subject to the constraint \eqref{eq:ParticleNumber_mod}. Formally this corresponds to finding the minimum of
$E[\{c_n,\bar c_n\}]- \mu \sum_{n=0}^{N_v} \bar c_n c_n$,
with the Lagrange multiplier $\mu$ being the chemical potential. Extremizing with respect to $\bar c_n$ yields the set of equations
\begin{eqnarray}
\label{eq:chemPot}
\frac{\delta E[\{c_n,\bar c_n\}]}{\delta \bar c_n} = \mu  c_n,
\end{eqnarray}
for $n=0,1, \dots, N_v$. By multiplying this equation by $\bar c_n$ and summing over $n$ we obtain an explicit expression for the chemical potential
\begin{eqnarray}
\label{eq:chemPotExplicit}
\mu =  \frac{2 E[\{c_n,\bar c_n\}]}{N},
\end{eqnarray}
where we used eq. \eqref{eq:ParticleNumber_mod} and the fact that $E[\{c_n,\bar c_n\}]$ is a homogeneous polynomial of degree two in the $c_n$'s.   
The solution to the system of equations \eqref{eq:chemPot} is most conveniently found by numerically time-evolving the system of ODEs \eqref{eq:GPforCoeffs} in imaginary time $t\rightarrow i \tau$. The right hand side of \eqref{eq:GPforCoeffs} being the energy gradient in the space of the $\{c_n\}$, this iterative optimization procedure is a gradient descent algorithm. We start out with a random seed of $\{c_n\}$'s and repeat a cycle of thousands of imaginary time-evolution steps followed by normalization of the $c_n$ to satisfy the constraint \eqref{eq:ParticleNumber_mod}. We stop the iteration when the energy decrement $\Delta E$ per cycle is negligible, which we define as $\Delta E/E < 10^{-12}$. 

This algorithm yields the ground state solution $\{c_n^{(0)}\}$ that minimizes the energy for a given coupling strength $g'$ and has the boson number $N$. Then by virtue of eqs. \eqref{eq:GPforCoeffs} and \eqref{eq:chemPot} the time-evolution of the ground state is given by
\begin{eqnarray}
\label{eq:grdStatetimedep}
c_n^{(0)}(t)= c_n^{(0)} e^{-i\mu t}.
\end{eqnarray}

Once the ground state solution $\{c_n^{(0)}\}$ is found, we can visualize it as follows: 
Since $\psi(z,t)=\sum _{n = 0} ^{N_v} c^{(0)}_n(t) \psi^{\text{LLL}}_n(z)=\left[\sum _{n = 0} ^{N_v} c_n^{(0)}(t) \mathcal  N_n z^{n}\right] e^{-|z|^2 /4 l_B^2}$, the expression in the bracket is a polynomial in $z$ of degree $N_v$. The $N_v$ zeros of this polynomial determine the locations of the superfluid's vortices. Denoting the roots of the polynomial by $z_i$, we can write $\psi(z,t)$ as a product of linear factors
\begin{eqnarray}
\label{eq:linearFactors}
\psi(z,t)= c^{(0)}_{N_v}(t)\prod_{n = 1} ^{N_v} \left[z-z_i\right] e^{-|z|^2 /4 l_B^2}.
\end{eqnarray}
Carrying out this factorization numerically, we can transform from the ground state configuration $\{c_n^{(0)}\}$ to the ground state configuration of the vortex locations $\{z_i\}$. The sole time-dependence of the solution \eqref{eq:grdStatetimedep} is contained in the overall phase factor. Since this factor does not affect the location of the zeroes of $\psi(z,t)$, we see that the vortices in the ground state are static. 
In Fig. \ref{fig:VortexLattice} we show the normalized bosonic density $|\psi(x,y)|^2/\max_{x,y} |\psi(x,y)|^2$ obtained by our numerical procedure for $N_v=40, 100, 300, 400$. Clearly the vortices are arranged into a regular pattern. The vortex lattice in a finite disk geometry investigated in this paper has some peculiarities near its edge.  For a sufficiently large droplet, deep inside the bulk the vortex arrangement is a triangular crystal, while close to the boundary the vortices form a nearly-equidistant circular pattern, which is not compatible with the triangular lattice. As a consequence, the finite vortex lattice is frustrated. This fact is particularly important for the study of surface excitations, since the collective motion in this case takes place almost exclusively on the boundary, where any description that employs a triangular vortex lattice is inadequate. 

We have solved eq. \eqref{eq:chemPot} for a fixed value of $N$. We can obtain the solution for any other particle number $\tilde N=\alpha N$, where $\alpha$ is a positive constant, by rescaling the solution \eqref{eq:grdStatetimedep}. It follows from eqs. \eqref{eq:ParticleNumber_mod} and \eqref{eq:chemPot} that the set $\{ \sqrt{\alpha} c_n^{(0)}\}$ is the ground state with $\tilde N$ particles at the same value of $g'$. It is found from eq. $\eqref{eq:chemPotExplicit}$ that the chemical potential changes by the factor $\alpha$. Of course one should keep in mind that our analysis is only valid in the limit of large filling fractions $N/N_v$, since our starting point is the Gross-Pitaevskii equation, which can only account for vortex crystals that reside in the mean-field regime.

\begin{figure*}[t] 
\centering{}\includegraphics[width= 17.0 cm]{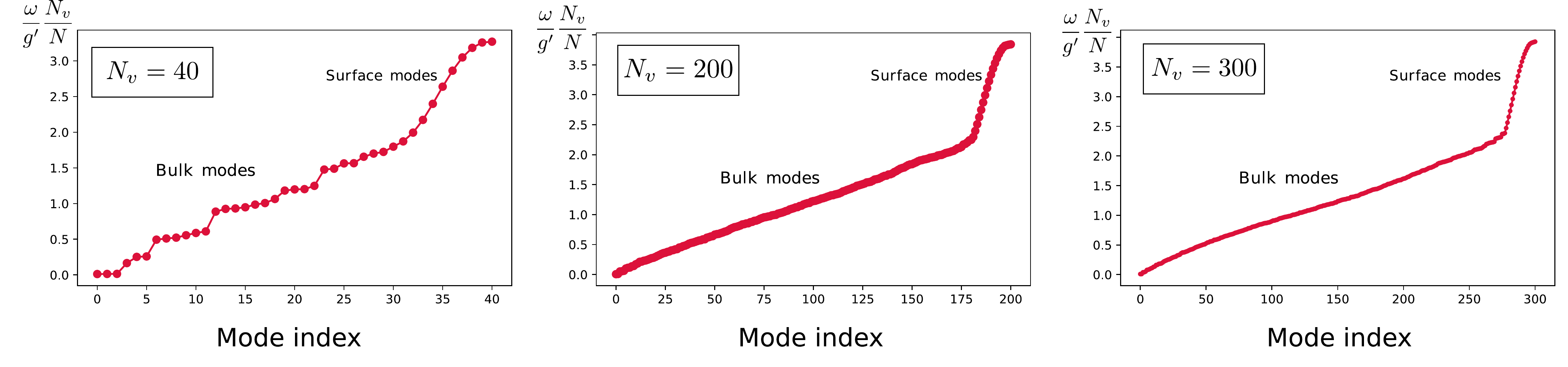}\protect\caption{\label{fig:FrequencySpectrum} Frequency spectrum of a crystal with $N_v = 40, 200$ and $300$ vortices obtained by diagonalizing the block matrix in eq. \eqref{eq:bdgEV}.  The frequencies are labeled, starting with index $0$, from smallest to largest frequency. The appearance of the spectrum has two qualitatively distinct parts. By visualizing the eigenfunctions we find that these two groups are the bulk and surface modes, respectively.}
\end{figure*}

\subsection{Collective small-amplitude oscillations}
\label{sec:CollectiveOsc}

Having found the ground state, we can now study the small-amplitude oscillations around this background by following the usual Bogoliubov-de Gennes procedure \cite{pitaevskii2016bose}. We assume now that the vortex crystal is perturbed slightly such that 
\begin{eqnarray}
\label{eq:BdGAnsatz}
c_n(t) = \left[c_n^{(0)} + \delta c_n(t)\right] e^{-i\mu t}
\end{eqnarray}
with small $\delta c_n(t)$ and work out the equation governing its dynamics. From here on we use a compact notation by forming the $(N_v+1)$-dimensional  vector $\bm{c}=(c_{0},\dots,c_{N_{v}})$. The equation for $\delta \bm{c}$ is found from \eqref{eq:GPforCoeffs} by linearization
\begin{eqnarray}
\label{eq:deltacnEOM}
i\frac{d}{dt}\delta\bm{c}=M^{1}\delta\bm{c}+M^{2}\delta\bar{\bm{c}},
\end{eqnarray}
with matrices $M^1$ and $M^2$ given by
\begin{equation}
\begin{split}
M_{nl}^{1}&=	4g'\sum_{s=0}^{2N_{v}}p_{l}^{s}p_{n}^{s}c_{s-l}^{(0)}\bar{c}_{s-n}^{(0)}-\mu \delta_{nl},\\
M_{nl}^{2}&=	2g'p_{n}^{n+l}\sum_{m}p_{m}^{n+l}c_{m}^{(0)}c_{n+l-m}^{(0)}.
\end{split}
\end{equation}
These $(N_v+1)\times(N_v+1)$ dimensional matrices only depend on ${c_n^{(0)}}$ and are therefore entirely determined by the vortex crystal ground state. The matrix $M^1$ is hermitian, while $M^2$ is a Hankel matrix.   As discussed at the end of the previous section, the particle number increases by a factor $\alpha$ if we multiply all the $c_n^0$ by $\sqrt{\alpha}$. Thus the entries of the matrices $M$  are all proportional to the particle number.  As a consequence, all oscillation frequencies scale linearly with $N$. Therefore in the following we plot  the frequency spectra after rescaling by a factor $N_v/N$.

To find the modes of oscillation, we first note that the equation of motion \eqref{eq:deltacnEOM} connects $\delta \bm{c}$  to $\delta \bar{\bm{c}}$, thus a single frequency ansatz with $e^{i \omega t}$ cannot solve eq. \eqref{eq:deltacnEOM}. Instead, we make an ansatz that also includes oscillations with the negative frequency according to
\begin{eqnarray}
\label{eq:bdgAnsatz}
\delta\bm{c}=\delta\bm{u}e^{i\omega t}+\delta\bar{\bm{v}}e^{-i\omega t} 
\end{eqnarray}
with some complex vectors $\delta\bm{u}$ and  $\delta\bm{v}$.  Insertion into the equation of motion \eqref{eq:deltacnEOM} results in an $(2N_v+2)\times (2N_v+2)$ eigenvalue problem that we write compactly in a block-matrix form as
\begin{eqnarray}
\label{eq:bdgEV}
\left(\begin{array}{cc}
-M^{1} & -M^{2}\\
\bar{M}^{2} & \bar{M}^{1}
\end{array}\right)\left(\begin{array}{c}
\delta\bm{u}\\
\delta\bm{v}
\end{array}\right)=\omega\left(\begin{array}{c}
\delta\bm{u}\\
\delta\bm{v}
\end{array}\right).
\end{eqnarray}

We note that the simplicity of the form of the Bogoliubov-de Gennes equations \eqref{eq:bdgEV} stems from making use of the LLL approximation \eqref{psiLLL} for the condensate.
The eigenvalue spectrum yields the oscillation frequencies $\omega$, while $\delta \bm u$ and $\delta \bm v$ describe the eigenmode. As usual, the doubling of the degrees of freedom by the introduction of the vector $(\delta\bm{u}, \delta\bm{v})^T$ results in an apparent doubling of the collective modes. However, as we prove in App. \ref{AppB}, for every eigenmode $(\delta\bm{u}, \delta\bm{v})^T$ with frequency $\omega$, the vector $(\delta \bar{\bm{v}}, \delta \bar{\bm{u}})^T$ is an eigenmode of \eqref{eq:bdgEV} with frequency $-\omega$. Since upon insertion in \eqref{eq:bdgAnsatz} it yields the same $\delta \bm c$, from here on we disregard the negative frequency modes of the spectrum. Below we find numerically that the block matrix in \eqref{eq:bdgEV} has only real eigenvalues. 

Due to $U(1)$ phase and spatial rotational symmetries, spontaneously broken by the ground state, the linearized Gross-Pitaevskii equation \eqref{eq:deltacnEOM} has exact zero modes. First, notice that one can construct an exact zero-frequency mode of the full non-linear LLL Gross-Pitaevskii equation that rescales all the ground state coefficients $\{c_n^{(0)}\}$ by the same complex phase. Such rescaling corresponds to a global $U(1)$ symmetry transformation of the condensate wavefunction and thus leaves the vortex positions unmodified. The corresponding linearized eigenvector of this mode is
$\delta c_n = i \alpha c_n^{(0)}$,
where $\alpha$ is a real constant, see App. \ref{AppC} for the proof.
Another exact zero mode follows from the global rotation symmetry of our geometry. Any global rotation of the ground state around the center of the condensate yields another valid ground state. In App. \ref{AppC} we work out that an infinitesimal rotation by angle $\theta$ gives rise to a zero-mode of the linearized equation \eqref{eq:deltacnEOM} with 
$\delta c_n = i \theta n c_n^{(0)}$.
In a recent publication \cite{polkinghorne2021two} the presence of these two exact zero modes has been rigorously demonstrated under more general conditions (in particular, the LLL limit is not required for their presence). Our results are consistent with the findings of \cite{polkinghorne2021two} and in App. \ref{AppC} we provide an explicit proof of the existence of the two zero modes in the LLL formalism. 

Curiously, for certain vortex numbers, we observe extra modes with nearly zero frequency. First, we checked numerically that for all ground states with $N_v = 3, 4, \dots, 9$, there are only two zero modes, as expected. Starting with $N_v = 10$, however, we found an additional soft mode. For some values of vortices, such as $N_v = 38$ there are even two extra soft modes. However, for the larger lattices  $N_v > 80$ that we studied, we find that these extra soft modes occur less frequently. We therefore suspect that the occasional appearance of additional soft modes is tied to the presence of frustration effects that are quite pronounced for small lattice sizes. We defer a detailed investigation of these unexpected low-frequency modes to future work.


\section{From the bulk Ruderman mode to surface waves}

By  diagonalizing the BdG matrix in eq. \eqref{eq:bdgEV} we obtain the frequency spectrum for different values of $N_v$. Examples are shown in Fig. \ref{fig:FrequencySpectrum}, where we have numbered the modes from smallest (with index $0$) to the largest frequency. By applying the procedure outlined at the end of \ref{sec:vortexLatticeGS}, we can visualize the eigenmodes  \eqref{eq:BdGAnsatz}. We observe that the lower frequency modes are bulk excitations, where all the vortices in the crystal move to an appreciable degree. On the other hand, we find that the high-frequency modes in Fig. \ref{fig:FrequencySpectrum} are in fact surface waves, because the motion of the vortices takes place almost exclusively in the outermost layer of the crystal.

\begin{figure}[] 
\centering{}\includegraphics[width= 0.8\columnwidth]{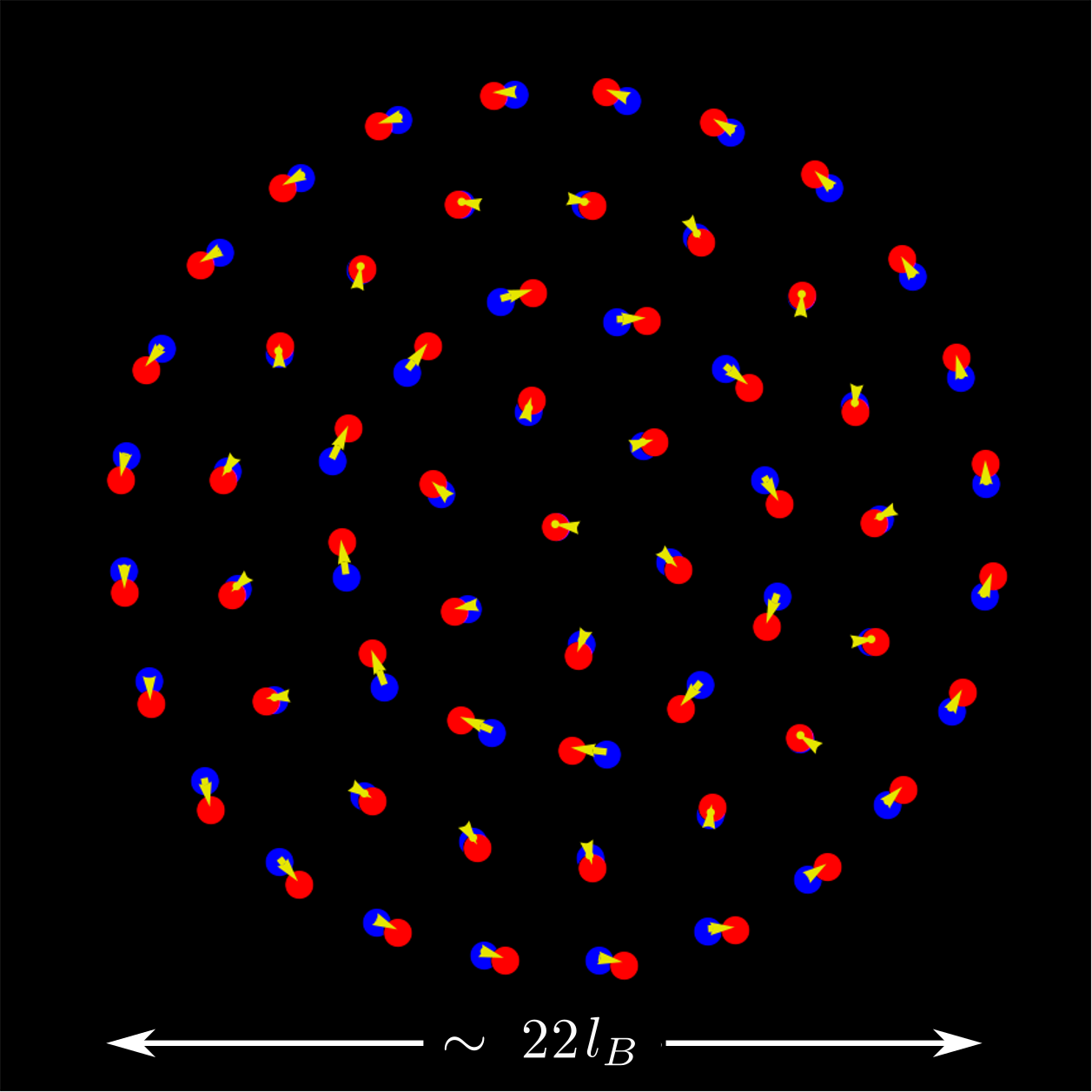}\protect\caption{\label{fig:TorsModes} Vortex displacement pattern of the low-energy torsional Ruderman mode for $N_v=60$. The blue dots show the undeformed lattice. The arrows are the displacement vectors to the new vortex positions (red). }
\end{figure}

We start our discussion with the investigation of the low-frequency bulk waves. In Fig. \ref{fig:TorsModes} the vortex deformation pattern of a low frequency mode with index $3$ is illustrated for $N_v = 60$ vortices.  This is a torsional mode, in which the vortices are moving predominantly azimuthally, see \cite{supmat} for a video of this type of motion. A mathematical treatment of torsional waves in vortex crystals of incompressible superfluids was first developed by Ruderman \cite{ruderman1970long} who attempted to explain slow oscillations in the period of pulsars after glitches, see reviews \cite{1987sonin, Sonin2016}. Here we adapt his line of reasoning to our model in order to understand the mode shown in Fig. \ref{fig:TorsModes}. Our starting point is an EFT for two-dimensional superfluid vortex crystals valid at small frequencies and long wavelengths, that some of us have worked out in \cite{effectivevortex,moroz2019bosonic,nguyen2020fracton}, see App. \ref{AppA} for a brief summary. Within the lowest-order derivative approximation, the vortex crystal is treated as an incompressible elastic medium, i.e.,  the displacement field $u^i$ can be written as the skew derivative $\epsilon^{ij}\partial_j \varphi$ of the potential function $\varphi$. In App. \ref{AppA} we derive that within the leading order EFT $\varphi$ satisfies the following equation 
\begin{eqnarray}
\label{eq:waveequation}
\partial^2_t \varphi - \frac{2C_2 \varepsilon''}{B_0^2} \nabla^4 \varphi = 0.
\end{eqnarray}
Here $C_2$ denotes the shear modulus of the vortex lattice, $B_0$ is the magnetic field and $\varepsilon''$ is the second derivative of the equation of state with respect to the density of bosons, see App. \ref{AppA} for details. The equation of motion \eqref{eq:waveequation} entails a quadratic dispersion relation $\omega \sim k^2$ valid at low momenta. We can readily solve this equation in the disk geometry using polar coordinates by noting that $J_n(k r) e^{i n \theta}$ is an eigenfunction of the Laplacian with eigenvalue $-k^2$. We make an ansatz
\begin{eqnarray}
\label{eq:phiAnsatz}
\varphi = \varphi_0 J_n(k r) e^{i n \theta} e^{i \omega t}
\end{eqnarray}
which is indeed a solution of eq. \eqref{eq:waveequation} with the dispersion relation 
\begin{eqnarray} \label{Tkfre}
\omega^2 = \frac{2C_2 \varepsilon''}{B_0^2} k^4.
\end{eqnarray}
The corresponding displacement field is given by 
\begin{eqnarray}
\label{eq:phiAnsatzDisplacement}
\bm u_n(k)  = 
\left(\begin{array}{c}
u_r\\
u_\theta
\end{array}\right)=
\varphi_0 
\left(\begin{array}{c}
\frac{i n}{r}  J_n(k r)\\
-k J_n'(k r)
\end{array}\right) e^{i n \theta} e^{i \omega t}.
\end{eqnarray}

\begin{figure}[t] 
\centering{}\includegraphics[width= 8.0 cm]{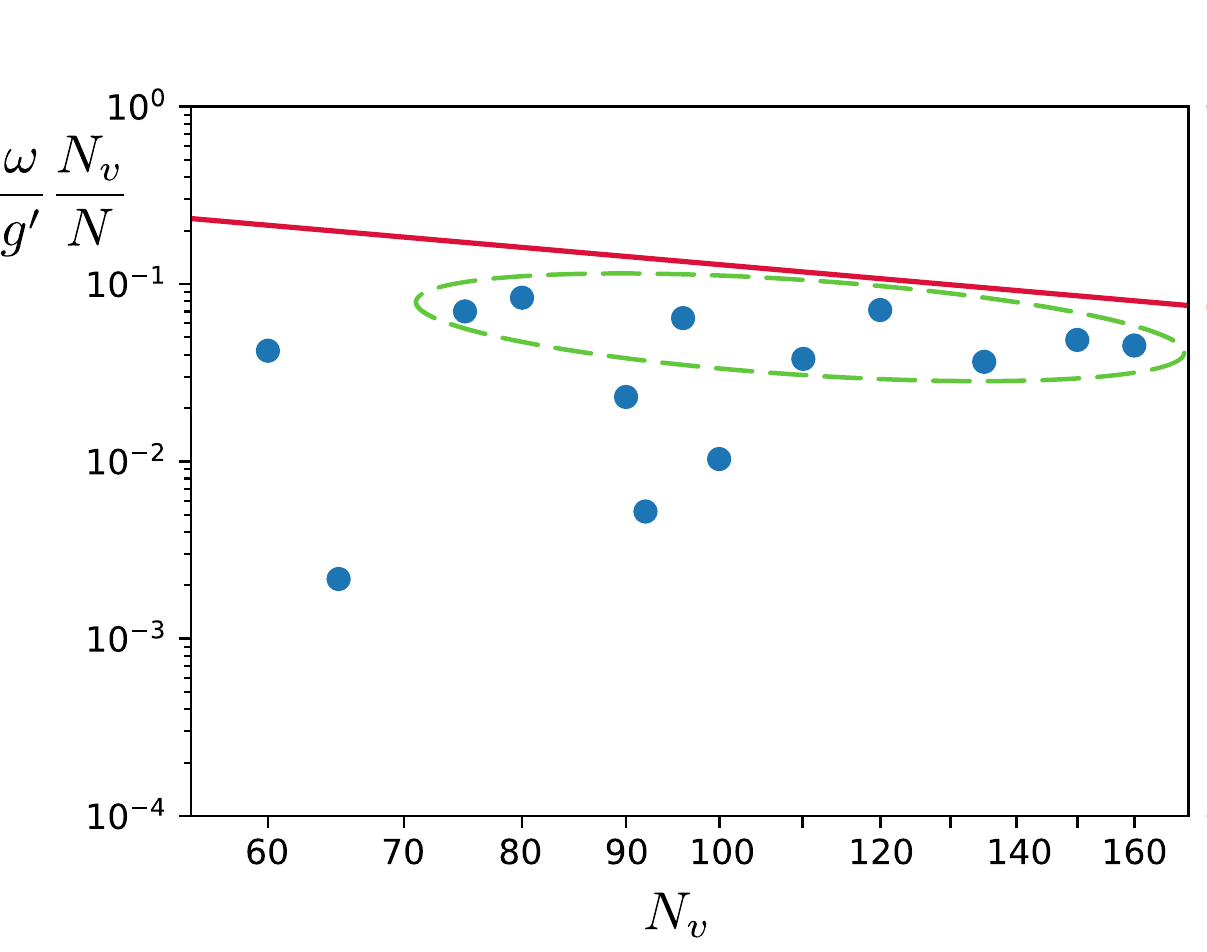}\protect\caption{\label{Fig:freqComparison} Comparison of the frequency of the lowest Ruderman mode obtained from the numerical solution (blue dots) and the analytical prediction from EFT (red line). For the system-sizes encircled in green the torsional mode frequency agrees to within a factor of $3$ with the estimate in eq. \eqref{Ruder}.}
\end{figure}

\begin{figure*}[t] 
\centering{}\includegraphics[width= 17.0 cm]{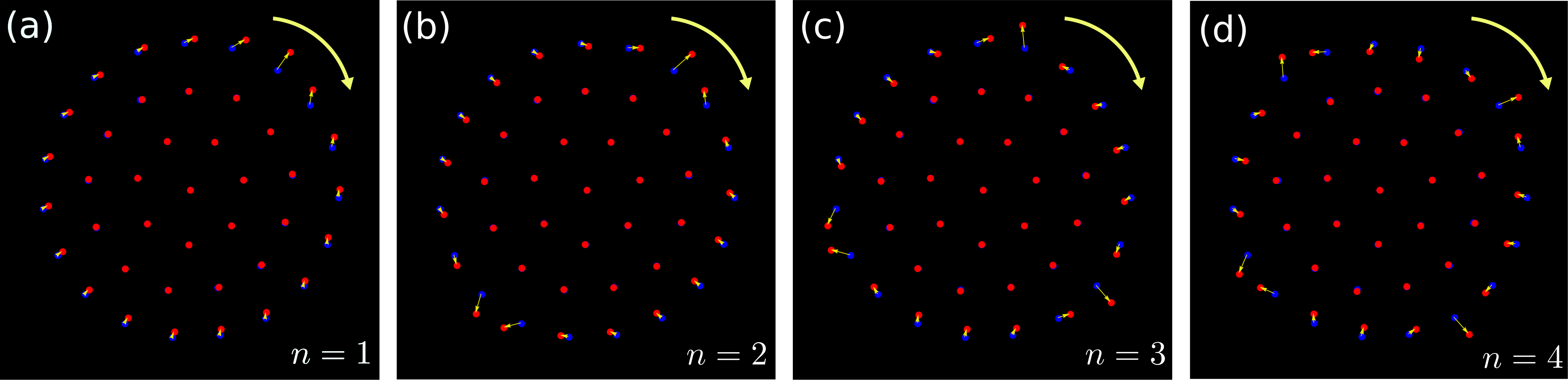}\protect\caption{\label{fig:SFModes} Vortex displacement patterns (a)-(d) for the four largest frequency surface modes $40, 39, 38, 37$ in a crystal with $N_v =40$ vortices. The blue points show the undistorted lattice, the yellow arrows are the displacement vectors pointing to the displaced vortices (red). The number $n$ of modulations in the displacement field of the outermost ring increases from $1$ to $4$. The oscillation pattern rotates here in the clockwise direction demonstrating the chiral nature of the surface waves. }
\end{figure*}

In general, the allowed values of the momentum $k$ are fixed by boundary conditions. Given the rotational symmetry of our geometry, we will impose that the mixed component of the stress tensor $T_{r\theta}$ vanishes at the boundary located at radius $R$. This implies the absence of any radial flux of azimuthal momentum through the boundary and thus the conservation of total angular momentum. As explained in App. \ref{AppA}, within the leading order EFT, the stress tensor is simply $T_{ij}=4C_2 u_{ij}$, where $u_{ij}$ denotes the strain tensor. Thus we end up with the Ruderman boundary condition expressed in polar coordinates as
\begin{equation}
   \Big( \frac{\partial u_{\theta}}{\partial r}- \frac{u_\theta}{r}+\frac 1 r \frac{\partial u_r}{\partial \theta} \Big)\Big|_{r=R}=0.
\end{equation}
Substituting now the expressions \eqref{eq:phiAnsatzDisplacement} with $n=0$ into this equation, we obtain the condition $J_2(k R)=0$.
This gives rise to the following quantization of momenta
\begin{equation}
k_s = \frac{j_{2,s}}{R},
\end{equation}
where $j_{2,s}$ is the $s$-th root of the Bessel function $J_2(x)$. The lowest Ruderman mode has a wavenumber that is determined by the first root $j_{2,1}\approx5.136$
\begin{equation}
k_\text{R} = \frac{5.136}{R}
\end{equation}
with frequency \eqref{Tkfre} 
\begin{equation} \label{Ruder}
\omega_\text{R} = \frac{\sqrt{2 C_2 \varepsilon''}}{B_0} k^2 \sim 13 \sqrt{\beta} \frac{N}{N_v^2} g'.
\end{equation}
In the last step we used $\varepsilon''= 
\beta g$, where $\beta$ is the Abrikosov parameter \cite{Abrikosov1957} of the vortex lattice. In the case of the infinite triangular lattice this value is known to be $\beta = 1.1596$. In our case the ground state is triangular at the center and becomes circular near the boundary. Thus in the present problem $\beta$ should be strictly speaking position-dependent. However, for the purpose of our order-of-magnitude estimate in eq. \eqref{Ruder}, we treat it as a constant and set $\beta = 1$ throughout. 
In eq. \eqref{Ruder} we also used the LLL expression for the shear modulus $C_2$ derived for the vortex lattice in \cite{sinova2002quantum, sonin2005ground} 
\begin{equation}
C_2 = 0.119 g n^2,    
\end{equation}
where $n$ is the coarse-grained boson density.

In Fig. \ref{Fig:freqComparison} we compare the analytical estimate eq. \eqref{Ruder} to the frequency of the Ruderman mode for various vortex numbers $N_v$. We obtained these frequencies by visualizing for each $N_v$ the numerically obtained displacement fields and identifying a low-lying mode that has close agreement with the displacement pattern found in eq. \eqref{eq:phiAnsatzDisplacement}. An example of such a visualization is shown in Fig. \ref{fig:TorsModes} for the case of $N_v = 60$. 

We observe that for many vortex numbers $N_v$ (those circled in green) there is a reasonable agreement, to within a factor $3$, with the estimated frequency eq. \eqref{Ruder}.  Yet there are also systems, particularly at smaller values of $N_v$, where this estimate fails dramatically. In these systems we observe that the Ruderman mode  becomes a soft-mode, already mentioned in the previous section, with a frequency that is smaller than the estimate \eqref{Ruder} by up to factors of $900$.  We suspect that the underlying cause of this disagreement is closely tied to lattice frustration effects, that are particularly pronounced for small systems, which are not captured by the EFT. We defer the detailed investigation of these special system sizes to future work.

Next we turn to a discussion of the surface waves. Using the leading order version of the same EFT that we analyzed above, we prove in App. \ref{AppA}, that the crystal cannot support Rayleigh waves in the low-frequency and large-wavelength realm. The basic reason for this is the incompressible nature of the vortex crystal in this regime. 

While low-frequency surface modes are absent, the numerical diagonalization of the microscopic BdG equations in the disk geometry reveals the presence of surface waves near the top of the frequency spectrum, see Fig. \ref{fig:SFModes} for snapshots and \cite{supmat} for videos. We found that for these modes the bulk of the vortex crystal is almost at rest, while only the outermost layer of the crystal is in motion. The vortices in that layer move on nearly elliptical  orbits around their resting positions. The shape of the displacement field rotates as a whole in a clockwise direction. The latter is determined solely by the sign of the underlying external magnetic field. A switch of the sign reverses the propagation direction of the surface waves by converting the holomorphic solutions \eqref{eq:basisfunctions} into antiholomorphic LLL functions. 

\begin{figure}[t] 
\centering{}\includegraphics[width= 8.0 cm]{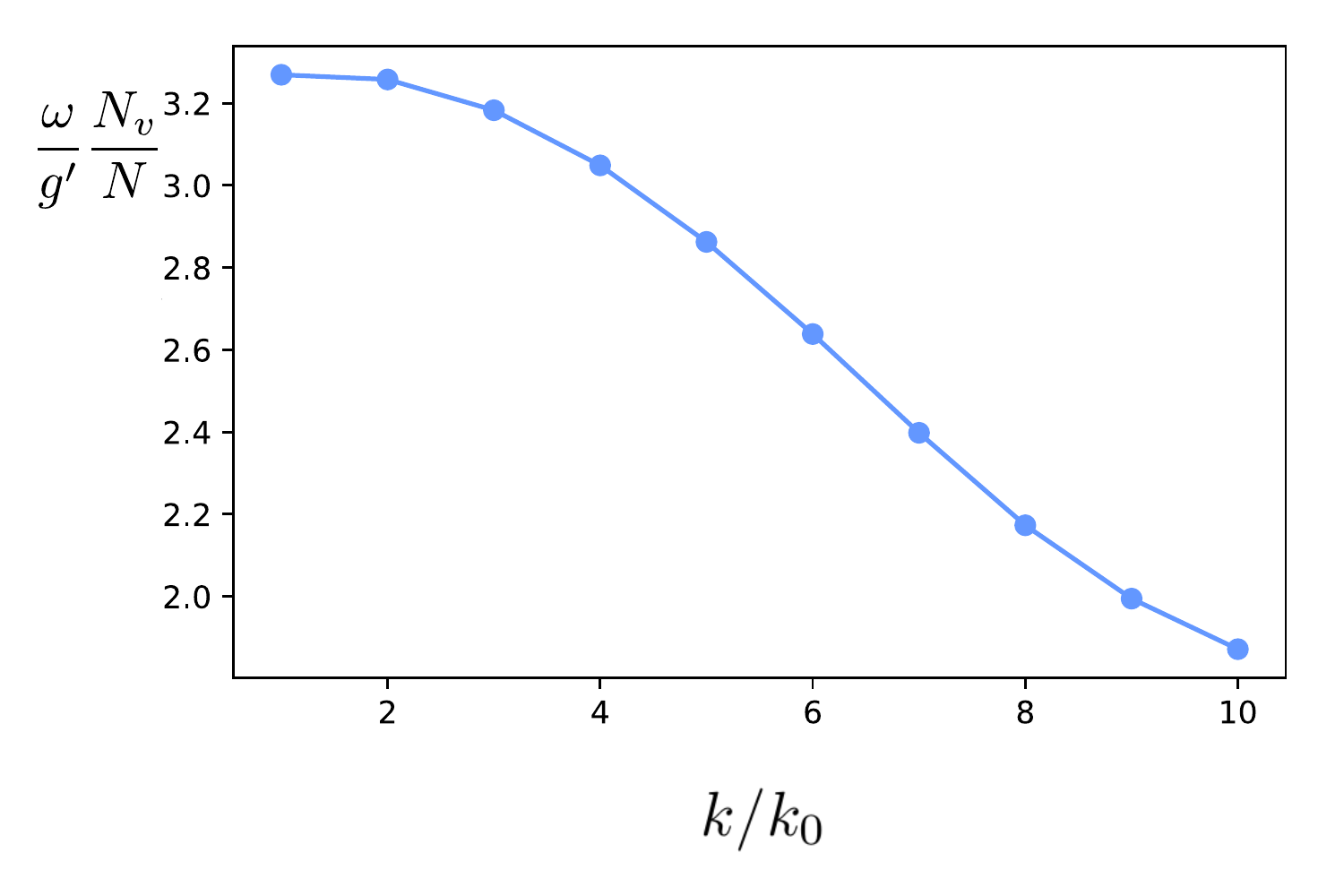}\protect\caption{\label{fig:DispEdge} Dispersion of the surface modes for $N_v = 40$. The horizontal axis is the wavenumber in units of $k_0 = 1/R$, where $R$ is the radius of the circular droplet.}
\end{figure}

The frequency of the surface waves is fixed by the coupling $g'$, which is the only energy scale in the LLL approximation. More precisely, the near-identical form of the spectra for $N_v \gtrsim 100$, clearly seen in Fig. \ref{fig:FrequencySpectrum}, implies that the surface wave frequencies scale with $N$ and $N_v$ as $\omega  \sim 4 g' N/N_v$. The wavelength of the highest mode excitation is equal to the circumference $C=2\pi R$ of the droplet. The wavelength decreases to $C/2$, $C/3$, etc. as we consider eigenmodes with progressively decreasing index, see Fig. \ref{fig:SFModes}. The edge-wave dispersion is shown in Fig. \ref{fig:DispEdge}. It has a peculiar feature that the frequency decreases with the increasing wavenumber.

\section{Summary and outlook}
After demonstrating that a two-dimensional superfluid vortex crystal does not support low-frequency Rayleigh edge waves, we have undertaken a microscopic study based on the investigation of normal modes of the Gross-Pitaevskii equation in the disk geometry in the LLL approximation. 
Numerically, we revealed that chiral surface waves with a peculiar dispersion relation emerge in vortex crystals at high frequencies. Moreover, we investigated carefully the
low-frequency torsional Ruderman bulk excitation. It is reassuring that for many values of $N_v$ our numerical microscopic results have reasonable agreement with the predictions of the leading-order low-energy effective theory of quadratically-dispersing Tkachenko waves \cite{redundancies, effectivevortex}. As noticed above, for certain values of $N_v$, especially those smaller than $80$, the torsional modes become soft. The origin of these additional soft modes presents a puzzle, which may have its origin in the competition between the triangular-lattice bulk and the circular boundary of the finite vortex crystal. We leave this as an open question for future investigations. 

The physics of the high-frequency chiral surface waves deserves further study. Can one write down an effective edge theory of the outer layer that captures the essential features of the waves? How sensitive are these modes to the imposed boundary conditions? What happens to them as one goes beyond the LLL regime? We believe that these questions are fruitful directions for future investigations.



\begin{acknowledgements}
We acknowledge useful discussions with Nigel Cooper and Edouard B. Sonin.  Our work is funded by the Deutsche Forschungsgemeinschaft (DFG, German Research Foundation) under Emmy Noether Programme grant no.~MO 3013/1-1 and under Germany's Excellence Strategy - EXC-2111 - 390814868. This work is supported by
Vetenskapsr\aa det (grant number 2021-03685).

\end{acknowledgements}

\begin{widetext}

\appendix
\section{Absence of Rayleigh waves in leading order low-energy effective theory of vortex lattices}
\label{AppA}
Here we investigate the Rayleigh problem within the leading order vortex lattice effective theory developed in \cite{effectivevortex, moroz2019bosonic}. At low frequencies and long wavelengths the emergent physics is governed by the intertwined dynamics  of coarse-grained elasticity and superfluidity. The effective theory can be systematically organized in a derivative expansion, whose leading order quadratic Lagrangian reads
\begin{equation} \label{LOEFT}
    \mathcal{L}^{(2)}=-\underbrace{\frac{B_{0} n_{0}}{2} \epsilon_{i j} u^{i} \dot{u}^{j}-\mathcal{E}_{\mathrm{el}}^{(2)}(\partial u)}_{\mathcal{L}_{\mathrm{el}}^{(2)}}+B_{0} e_{i} u^{i}-\frac{\varepsilon^{\prime \prime}}{2} \delta b^{2}.
\end{equation}
Here the physical degrees of freedom are elastic lattice displacement vectors $u_i$ and emergent $u(1)$ electric and magnetic fields $e_i$ and $\delta b$. The dual magnetic field $\delta b$ is measured with respect to a finite background which is fixed by the superfluid background density $n_0$.  Via the boson-vortex duality \cite{peskin1978, DasguptaHalperin}, the dual gauge fields encode two-dimensional coarse-grained superfluid degrees of freedom. The parameter $B_0$ denotes the strength of a background effective magnetic field experienced by bosons, in neutral superfluids under rotation with the angular frequency $\Omega$, one has $B_0=2m \Omega$, where $m$ is the mass of the elementary boson. Triangular symmetry of the vortex crystal fixes the form of the elastic energy density to take a simple form
\begin{equation} 
    \mathcal{E}_{\mathrm{el}}\left(u_{i j}\right)=2 C_{1} u_{k k}^{2}+2 C_{2} \tilde{u}_{i j}^{2}
\end{equation}
with $\tilde{u}_{i j} \equiv u_{i j}-\left(u_{k k} \delta_{i j}\right) / 2$ being the traceless part of the symmetric strain tensor $u_{ij}=(\partial_i u_j+\partial_j u_i)/2$; $C_1$ and $C_2$ are the compressional and shear elastic moduli, respectively. Notably, at leading order the electromagnetic part contains only the magnetic term $\sim \delta b^2$ with a prefactor fixed by the curvature $\varepsilon''=d^2 \varepsilon/db^2$ evaluated at the minimum of the superfluid equation of state $\varepsilon(b)$. In contrast to a non-rotating superfluid, the leading order theory does not contain the dynamical electric term $\sim e_i^2$ because it is of the next-to-leading order in the derivative expansion \cite{effectivevortex}.

In the first step, we express the dual electric and magnetic fields in terms of the gauge potentials, namely $\delta b=\epsilon_{ij}\partial_i a_j$ and $e_i=\partial_t a_i - \partial_i a_t$, and perform the quadratic functional integral over dual gauge fluctuations $a_\mu$. Since the electric term $\sim e_i^2$ is absent in the leading order theory, only the spatial part of the gauge field appears quadratically in the Lagrangian. If one fixes the gauge as
$
    \mathcal{L}^{(2)}\to \mathcal{L}^{(2)}-\left(\partial^i a_i\right)^2/(2\xi) 
$, the Gaussian integration over $a_i$ results in a non-local elastic action that most conveniently can be written in momentum space \cite{ClaudioPhD}. The complete action it
\begin{equation}     \label{Sint} 
S_{\mathrm{eff}}\left[u^{i}, a_t\right]= \int dt d\mathbf{x} \left(\mathcal{L}_{\mathrm{el}}^{(2)}+ B_0 a_t \partial_i u^i\right)  + \frac{B_0^2}{2 \varepsilon''} \int  \frac{d t d \mathbf{k}}{(2\pi)^2} \left(\frac{\dot{u}_{\mathbf{-k}}^{i} \dot{u}_{\mathbf{k}}^{i}}{k^{2}}-\dot{u}_{\mathbf{-k}}^{i} \frac{(1-\xi)k_i k_j}{k^{4}} \dot{u}^{j}_{\mathbf{k}} \right),
\end{equation}
where overdot denotes the temporal derivative.
Since the emergent Newtonian term $\sim (\dot u^i)^2$ has a momentum-dependent prefactor that diverges at low momenta as $1/k^2$, within the derivative expansion it is as relevant as the Lorentz term (hidden in $\mathcal{L}_{\mathrm{el}}^{(2)}$) that contains only one temporal derivative.

The temporal gauge field $a_t$ appears only linearly in the theory \eqref{Sint} and functional integration over $a_t$ results in the Gauss constraint $\partial_i u^i=0$. We thus observe that at leading order the vortex crystal is incompressible! The constraint can be solved by introducing a potential field $\varphi$, namely $u^i=\epsilon^{ij}\partial_j \varphi$. In terms of $\varphi$, the theory becomes local with the following Lagrangian  \cite{redundancies}
\begin{equation} \label{Lvarphi}
    \mathcal{L}_{\varphi}= \frac{B_0^2}{2 \varepsilon''} \left[ (\partial_t \varphi)^2- \frac{2C_2 \varepsilon ''}{B_0^2} (\Delta  \varphi)^2 \right].
\end{equation}
The leading-order theory encodes purely transverse oscillations with the quadratic dispersion $\omega^2= 2C_2 \varepsilon'' k^4/B_0^2$. One must emphasize that the apparent time-reversal symmetry of the Lagrangian \eqref{Lvarphi} is completely accidental. It is known that next-to-leading order terms modify the polarization of the low-frequency Tkachenko wave by generating a small out-of-phase longitudinal component. The resulting elliptical chiral polarization of the Tkachenko wave explicitly indicates time-reversal symmetry breaking in vortex crystals.  

After this brief discussion of the salient properties of the leading-order effective theory of vortex crystals in two dimensions, we consider within this framework the surface elastic problem of Rayleigh \cite{Rayleigh1885}. In ordinary crystals, assuming no-stress boundary condition, he discovered linearly dispersing surface  waves localized close to an infinite straight boundary. Given that in the leading order theory  the vortex crystal is incompressible and supports quadratically dispersing transverse waves, one can anticipate that at this level of approximation one cannot verbatim repeat the arguments of Rayleigh who satisfied the no-stress boundary condition by superposing the longitudinal and transverse solutions of the elastic equations of motions.

We now investigate in detail the Rayleigh problem for the theory \eqref{Lvarphi}. Assuming the vortex crystal occupies the lower half-plane with a boundary at $y=0$, we start from the ansatz $\varphi(t, x, y)= e^{i(k x - \omega t)+\kappa y}$. We look for solutions oscillating in time and $x$-direction, moreover we are interested in waves that are exponentially-localized close to the boundary implying $\kappa>0$. By substituting this ansatz into the equations of motion that follow from the Lagrangian \eqref{Lvarphi}, we express the inverse of the dacay length in terms of $\omega$ and $k$
\begin{equation}
    \kappa_{\pm}(\omega, k)=\sqrt{k^2 \pm \frac{B_0}{\sqrt{2C_2 \varepsilon ''}} \omega}.
\end{equation}
The general solution with fixed $\omega$ and $k$ is a superposition of the two branches
\begin{equation} \label{Rs}
    \varphi= e^{i(k x - \omega t)} 
    \left( \varphi_+ e^{\kappa_+(\omega, k)y}+\varphi_- e^{\kappa_-(\omega, k)y} \right).
\end{equation}
Now, can one satisfy the stress-free boundary conditions $T_{xy}=T_{yy}=0$ by chosing appropriately $\varphi_+$ and $\varphi_-$? For the incompressible crystal, the Cauchy stress tensor is given by $T_{ij}=4 C_2 \tilde u_{ij}=4 C_2 u_{ij}= 4 C_2 \partial_{(i} \epsilon_{j) k}\partial_k \varphi$, where simple parentheses denote symmetrization of indices. As a result, the no-stress boundary conditions simplify to the conditions $\partial_x \partial_y \varphi=0$ and $(\partial_x^2 -\partial_y^2) \varphi=0$ imposed on the boundary. Substituting the solution \eqref{Rs} into these equations, one ends up with a homogeneous matrix equation for the coefficients $\varphi_+$ and $\varphi_-$
\begin{equation}
    \left(\begin{array}{cc}
ik \kappa_+ & ik \kappa_-\\
-k^2- \kappa_+^2 & -k^2- \kappa_-^2
\end{array}\right)\left(\begin{array}{c}
\varphi_+\\
\varphi_-
\end{array}\right)=0.
\end{equation}
By evaluating the determinant, we find that the linear system has a solution only for  $\omega(k)=0$. Since, however, that implies $\kappa_+=\kappa_-$ and $\varphi_+=-\varphi_-$, we find that the potential $\varphi$ in Eq. \eqref{Rs} vanishes and $\omega(k)=0$ is thus not a physical solution.

In conclusion, as verified above within the leading order effective theory \eqref{LOEFT}, two-dimensional superfluid vortex crystals do not support low-frequency and long-wavelength Rayleigh elastic surface waves.
\section{Properties of the BdG spectrum and its exact zero-frequency modes in the LLL regime}
\label{AppC}
\label{AppB}
We begin by showing that non-zero eigenfrequencies of the  Bogoliubov-de Gennes matrix always come in $\pm \omega$ pairs.
Let $(\delta\bm{u},\delta\bm{v})^{T}$ be a solution of the eigenvalue
problem with frequency $\omega$, then
\[
\left(\begin{array}{cc}
-M^{1} & -M^{2}\\
\bar{M}^{2} & \bar{M}^{1}
\end{array}\right)\left(\begin{array}{c}
\delta\bm{u}\\
\delta\bm{v}
\end{array}\right)=\omega\left(\begin{array}{c}
\delta\bm{u}\\
\delta\bm{v}
\end{array}\right).
\]
Taking the complex conjugate and switching the rows, we obtain:
\[
\left(\begin{array}{cc}
-M^{1} & -M^{2}\\
\bar{M}^{2} & \bar{M}^{1}
\end{array}\right)\left(\begin{array}{c}
\delta\bm{\bar{v}}\\
\delta\bm{\bar{u}}
\end{array}\right)=-\omega\left(\begin{array}{c}
\delta\bm{\bar{v}}\\
\delta\bm{\bar{u}}
\end{array}\right),
\]
where we used that $\omega$ is a real frequency. 
Thus we have proved that for an eigenvector $(\delta\bm{u},\delta\bm{v})^{T}$
with eigenfrequency $\omega$, there is another eigenstate $(\delta\bar{\bm{v}},\delta\bar{\bm{u}})^{T}$
with eigenfrequency $-\omega$. However, the latter is not a new physical solution, since it gives the same solution $\delta\bm{c}$ if inserted into eq. \eqref{eq:bdgAnsatz}.

In a rotation-invariant geometry, the Bogoliubov-de Gennes equations \eqref{eq:deltacnEOM} 
always have two exact zero-frequency solutions. In complete generality this has been noticed and proved by Polkinghorne and Simula in a recent paper \cite{polkinghorne2021two}.  Here we show how this emerges within the LLL formalism.

We first consider the global $U(1)$ transformation of the condensate. When we rescale the condensate $\psi$ by a phase $ \exp(i \alpha)$, where $\alpha$ is a real constant, we obtain another valid ground state of the Gross-Pitaevskii equation.  For small $\alpha$ this translates via eq. \eqref{psiLLL} into 
$\delta c_{n}=i\alpha c_{n}^{(0)}$.
We now explicitly show that this $\delta c_n$ satisfies eq. \eqref{eq:deltacnEOM} with eigenvalue zero. Upon insertion of $\delta c_n$ into eq. \eqref{eq:deltacnEOM} and cancellation of $i\alpha$ it remains to be shown that 
\begin{equation}
\label{eq:TrivialSol}
\sum_{l=0}^{N_{v}}M_{nl}^{1}c_{l}^{(0)}-\sum_{l=0}^{N_{v}}M_{nl}^{2}\bar{c_{l}}^{(0)}=0
\end{equation}
holds for all $n$. To prove this, we introduce the function
\[
\phi(k)=\sum_{m}p_{m}^{k}c_{m}^{(0)}c_{k-m}^{(0)},
\]
then the two terms in \eqref{eq:TrivialSol} can be compactly written
as 
\[
\sum_{l=0}^{N_{v}}M_{nl}^{1}c_{l}^{(0)}=4g'\sum_{s=0}^{2N_{v}}\phi(s)p_{n}^{s}\bar{c}_{s-n}^{(0)}-\mu c_{n}^{(0)}
\]
and
\[
\sum_{l=0}^{N_{v}}M_{nl}^{2}\bar{c_{l}}^{(0)}=2g'\sum_{l=0}^{N_{v}}\phi(n+l)p_{n}^{n+l}\bar{c_{l}}^{(0)}.
\]
Since the $c^{(0)}_n$ satisfy eqs. \eqref{eq:GPforCoeffs} and \eqref{eq:chemPot}, we have
\[
\mu c_{n}^{(0)}=2g'\sum_{s=0}^{2N_{v}}\phi(s)p_{n}^{s}\bar{c}_{s-n}.
\]
Combining all of these expressions, we see that the equation \eqref{eq:TrivialSol} holds and $\delta c_{n}=i\alpha c_{n}^{(0)}$ is in fact a solution of the Bogoliubov-de Gennes equations with zero frequency. 
Comparing with eq. \eqref{eq:linearFactors}, we see that this mode leaves the vortex positions unchanged and only modifies the overall phase factor of the Gross-Pitaevskii wavefunction.

Another zero frequency mode in the spectrum is generated by globally rotating all the vortices by a finite angle $\theta$. The new state remains a valid ground state of the Gross-Pitaevskii equation. In the language of vortices this amounts to rotating all of their positions $z_i$ by the factor $\exp(i\theta)$. In terms of $c_n$ such transformation corresponds, according to eqs. \eqref{psiLLL} and \eqref{eq:linearFactors}, to the transformation
\begin{equation}
\label{eq:cnRotSym}
c_n^{(0)} \rightarrow  e^{i n \theta} c_n^{(0)}.
\end{equation}
One can now explicitly check that the full equation of motion \eqref{eq:GPforCoeffs} is satisfied for the transformed $\{c_n^{(0)}\}$. For small $\theta$ the linearized transformation \eqref{eq:cnRotSym} becomes
\begin{equation}
\label{eq:DeltacnRotSym}
\delta c_n = i n \theta c_n^{(0)},
\end{equation}
which must satisfy the linearized equation \eqref{eq:deltacnEOM} by construction. An explicit proof along the lines of the previous demonstration can be straightforwardly constructed.

\section{External trapping potentials}
\label{app:Potential}
In the main text we assumed the presence of a real-space trapping potential that limits the occupied orbitals to $n\leq N_v$. Here we discuss potentials that realize this cutoff at $N_v$ in a soft and a hard manner. The idea is to let the number of included LLL orbitals be infinite and to introduce an external trapping potential that suppresses occupation of all orbitals with $n > N_v$.
\subsection{Soft-cutoff for LLL orbitals}
\label{app:SoftCutoff}
We consider a step-function radial potential with potential strength $V_{0}$ that confines the bosons 
\[
V_{\text{pot}}(r)=V_{0}\theta(r-R)=\begin{cases}
0 & r\leq R\\
V_{0} & r>R
\end{cases}
\]
where $\theta$ is the Heaviside-function. In the Gross-Pitaevskii energy functional this contributes a term
\[
E_{\text{pot}}=\int d^{2}\bm{r}\ \psi^{*}(\bm{r})V(\bm{r})\psi(\bm{r}).
\]
Employing the LLL limit we have 
\begin{equation}
\psi(z,t)=\sum_{n=0}^{\infty}c_{n}(t)\psi_{n}^{\text{LLL}}(z),\label{eq:psiExp}
\end{equation}
where we now include all LLL orbitals. Inserting this expansion into the expression for the potential energy, we obtain
\[
E_{\text{pot}} =2\pi V_{0}\sum_{n=0}^{\infty} \mathcal{N}_{n}^{2}\bar{c}_{n}(t)c_{n}(t)\int\limits _{R}^{\infty}dr\ r^{2n+1}e^{-r^{2}/2l_{B}^{2}},
\]
where we carried out the angular integration. 
Next we change the integration variable to $u=r^{2}/2l_{B}^{2}$
and obtain
\begin{align*}
E_{\text{pot}}  & =\frac{V_{0}}{n!}\sum_{n=0}^{\infty}\bar{c}_{n}(t)c_{n}(t)\int\limits _{R^{2}/2l_{B}^{2}}^{\infty}du\ u^{n}e^{-u}\\
 & =\sum_{n=0}^{\infty}V_{\text{pot}}^{n}(R^{2}/2l_{B}^{2})\bar{c}_{n}(t)c_{n}(t).
\end{align*}
Here
\[
V_{\text{pot}}^{n}(x)=\frac{V_{0}}{n!}\int\limits _{x}^{\infty}du\ u^{n}e^{-u}
\]
and the integral is the upper incomplete gamma function. The form of the $V_{\text{pot}}^{n}(x)$ is shown in Fig. \ref{Fig:HWPotential} for $x=100$. Clearly, only orbitals with $n\gtrsim x$ are appreciably suppressed.  In fact, this sigmoid function is well approximated by the hyperbolic tangent
\[
V_{\text{pot}}^{n}(x)\approx \frac{V_0}{2} \left( 1+ \tanh\left(\frac{n-x}{1.2\sqrt{x}}\right)\right),
\]
shown in red. Note that the crossover happens within a region of width of order $ \sqrt{x}$. 

\begin{figure}
\includegraphics[scale=0.81]{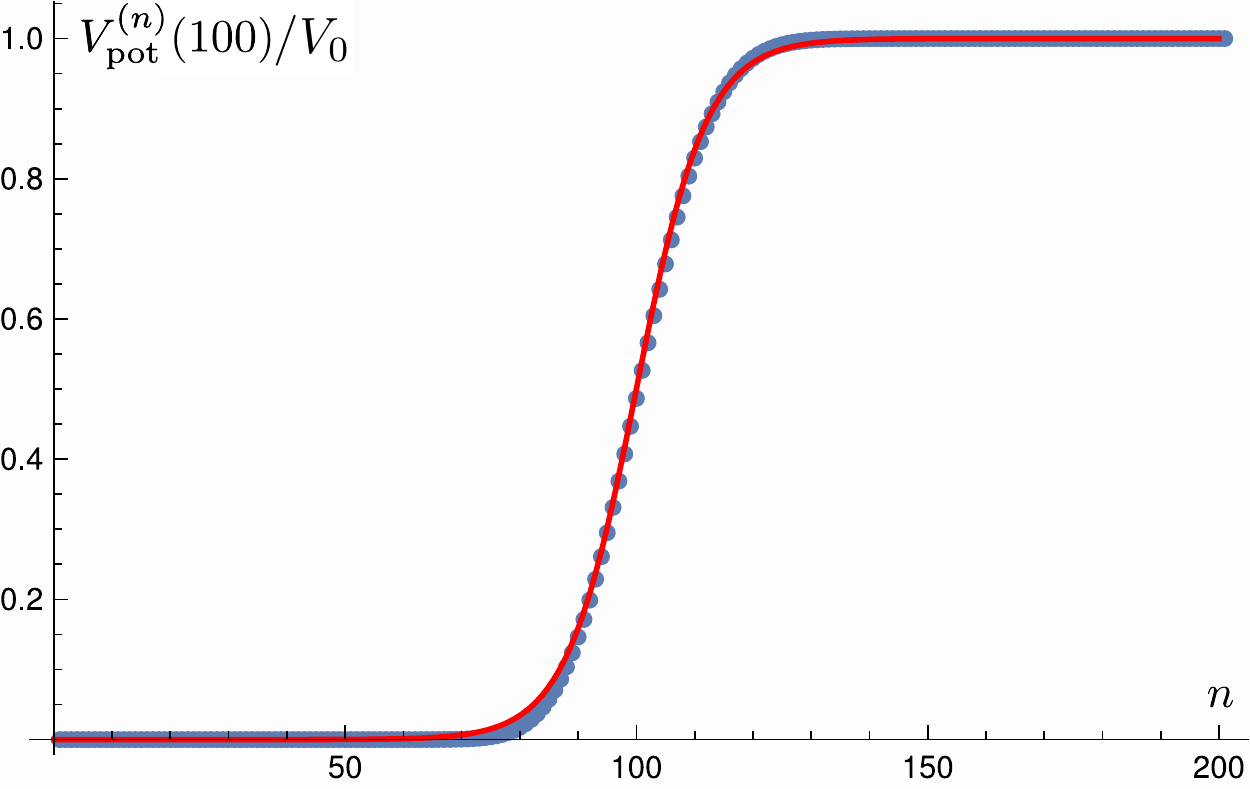}
\caption{\label{Fig:HWPotential} Plot of $V_{\text{pot}}^{(n)}(100)$ together with a sigmoid approximation (red), discussed in the text. }

\end{figure}

 Thus if we want to cut off $n$ at around $N_{v}$, then we just have to set $R^{2}/2l_{B}^{2}=N_{v}$, i.e. $R=\sqrt{2N_{v}}l_{B}$. 
 This radius agrees with the estimate for the radius of the crystal in the main text. If we now tune $V_{0}$ to be large compared with the interaction energy scale $g'$, but still much smaller than the cyclotron frequency (in order to remain within the LLL  approximation), this potential will strongly penalize terms with $n\gtrsim N_{v}$. The range of orbitals $n$ in which the transition from zero to strong penalty happens is roughly from $N_{v}-1.2\sqrt{N_{v}}$ to $N_{v}+1.2\sqrt{N_{v}}$. 

\subsection{Hard-cutoff for LLL orbitals}
\label{app:HardCutoff}
The previous calculation showed that a step-function potential  in real space results in a soft-cutoff for the LLL potential $V_{\text{pot}}^n$. Here we engineer a class of potentials that exactly realize a hard-cutoff for $V_{\text{pot}}^n$.

We saw above that within the LLL subspace a potential $V(r)$ leads to a contribution 
\[
\sum_{n}V_{\text {pot}}^{n}\bar{c}_{n}c_{n}
\]
with
\[
V_{\text {pot}}^{n}=\frac{2\pi}{2^{n+1}\pi n!l_{B}^{2n+2}}\int\limits _{0}^{\infty}dr\ V(r)r^{2n+1}e^{-r^{2}/2l_{B^{2}}}.
\]
Using the same substitution $u=r^{2}/2l_{B}^{2}$ and defining $W(r^{2}/2l_{B}^{2})  \equiv V(r)$, we can rewrite $V_{\text {pot}}^{n}$ as 
\begin{align*}
V_{\text {pot}}^{n} & =\frac{1}{n!}\int\limits _{0}^{\infty}du\ W(u)u^{n}e^{-u}.
\end{align*}
Now we make an ansatz for $W(u)$ in terms of Laguerre polynomials, because
they have the useful property
\[
\int\limits _{0}^{\infty}du\ L_{k}(u)u^{n}e^{-u}=(-1)^{k}n!{n \choose k}.
\]

\begin{figure}[t]
\includegraphics{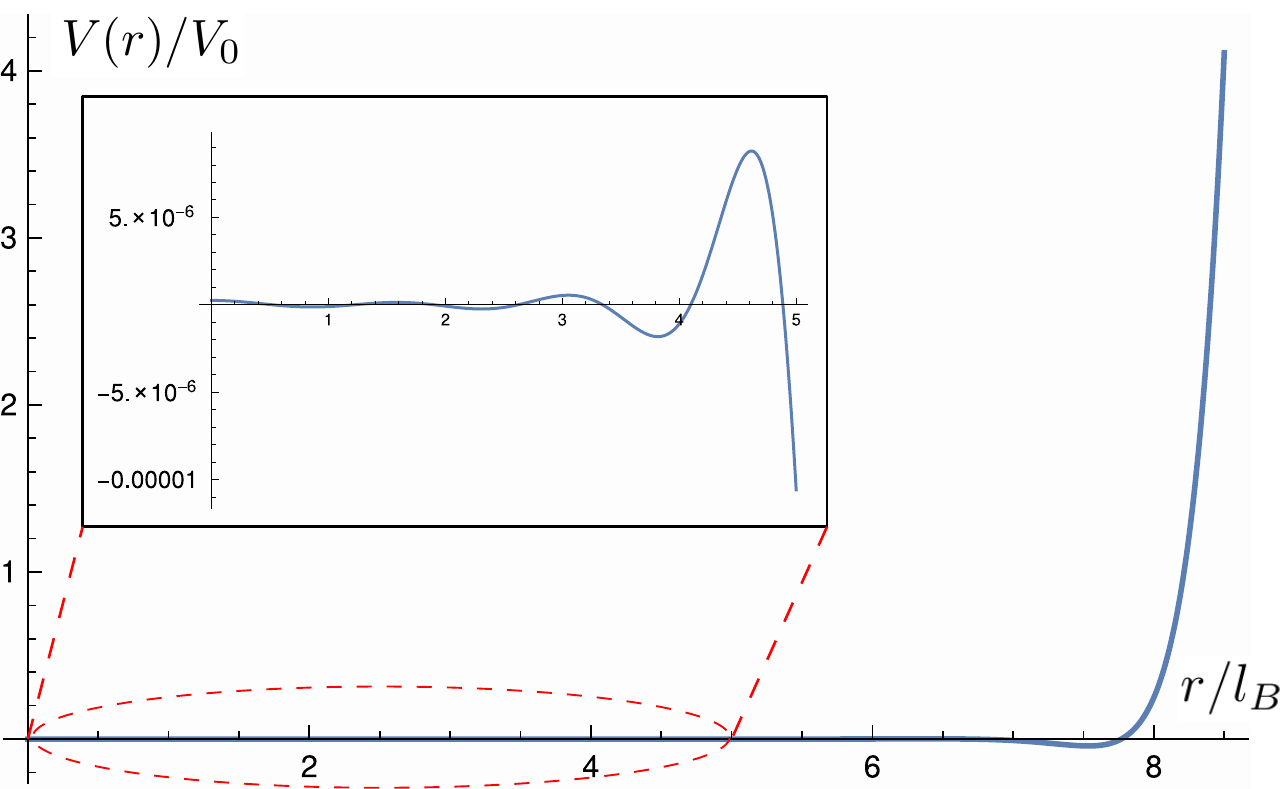}
\caption{\label{Fig:potentialV} Real-space form of the potential $V(r)$ for $N_{V}=10$ as a function of $r/l_B$.}
\end{figure}
Thus if $n<k$ this integral will vanish exactly. If we now pick
\[
W(u)=\sum_{k>N_{v}}\alpha_{k}L_{k}(u),
\]
with real coefficients $\alpha_k$, then automatically 
\[
V_{\text {pot}}^{n}=0 \text{ for } n\leq N_{v}.
\]
And for $n>N_{v}$ we have
\[
V_{\text {pot}}^{n}=\sum_{k>N_{v}}\alpha_{k}(-1)^{k}{n \choose k}.
\]
Clearly any choice of $\alpha_k$  with correctly alternating signs results  in a positive $V_\text{pot}^n$. We now specialize to the simple choice 
\[
\alpha_{k}=\frac{(-1)^{k}}{k!}V_{0},
\]
such that
\begin{equation}
V_{n}=\begin{cases}
0 & n\leq N_{v}\\
V_{0}\sum_{k=N_{v}}^{n}\frac{1}{k!}{n \choose k}. & n>N_{v}.
\end{cases}
\end{equation}
Thus for large $V_{0}$ the LLL orbitals with $n>N_{v}$ are penalized and therefore will remain vacant at low energies. The real-space form is given by
\begin{equation}
\label{eq:LagPot}
W(u)=V_{0}\sum_{k>N_{v}}\frac{(-1)^k}{k!}L_{k}(u).
\end{equation}
The potential converges to finite values for all $x$, since, up to a polynomial of degree $N_v$, it is equal to 
\[
\sum_{k=0}^{\infty}\frac{(-1)^{k}}{k!}L_{k}(u)=e^{-1}I_{0}(2\sqrt{u}),
\]
where $I_0$ is the zeroth-order modified Bessel function of the first kind.  
For $N_v=10$, the real-space form $V(r)$ is shown in the Fig. \ref{Fig:potentialV}. As expected it sharply rises around $R=\sqrt{2N_v} l_B$. Moreover, it exhibits oscillations for $r<R$ which are needed to ensure that $V_{\text {pot}}^{n}$ strictly vanishes for $n\leq N_v$.



\end{widetext}

\bibliography{biblio}

\end{document}